\def\k{\kappa}
\def\th{\theta}
\def\a{\alpha}
\def\beq{\begin{equation}}
\def\eeq{\end{equation}}

\documentclass{elsart}
\usepackage{epsfig}

\begin{document}
\begin{frontmatter}

\title{The energy spectrum of complex periodic potentials
of the Kronig-Penney type}

\author{H.~F.~Jones}
\address{Physics Department, Imperial College,  London SW7 2BZ, UK}

\begin{abstract}
We consider a complex periodic PT-symmetric potential of the
Kronig-Penney type, in order to elucidate the peculiar properties
found by Bender et al. for potentials of the form $V=i(\sin x)^{2N+1}$,
and in particular the absence of anti-periodic solutions. In this
model we show explicitly why these solutions disappear as soon
as $V^*(x)\neq V(x)$, and spell out the consequences for the form
of the dispersion relation.

\end{abstract}
\end{frontmatter}

In a recent paper Bender et al. \cite{BDM} showed that periodic
potentials which were complex but obeyed PT symmetry possessed
real band spectra, with, however, one striking difference from the
case of real periodic potentials, namely that there were no
antiperiodic solutions, i.e. Bloch waves with lattice wave vector
$k=(2n+1)\pi/a$. This result was obtained from detailed numerical
studies of potentials of the form $V(x)=i\sin^{2N+1}(x)$, and it
was found necessary to work to extremely high accuracy to detect
the absence of such solutions. In the present note we supplement
this work by an analytical solution to a complex, PT-symmetric
version of the Kronig-Penney model, which illustrates the
phenomenon very clearly. It therefore seems a generic property of
non-Hermitian, but PT symmetric potentials, although an analytic
proof is still not available.

In the standard Kronig-Penney model \cite{Kittel} the potential
consists of a periodic string of delta functions of the form \beq
V(x) = \a \sum_n \delta(x-na). \eeq The simplest way to find the
energy eigenvalues is the Floquet procedure, described in
\cite{BDM}, based on two solutions which in the region $0\le x <a$
are
\begin{equation}\label{Floquet}
 u_1(x) = \cos{\k x},\quad u_2=(1/\k)\sin{\k x}.
\end{equation}
The lattice wave vector $k$, which occurs in the phase ${\rm
e}^{ik a}$ of the Bloch wave function at $x=a_+$, is given by \beq
\cos ka = \half(u_1(a)+u_2'(a)). \eeq Crossing the delta
function at $x=a$, $u_2'$ has the discontinuity $\a u_2(a_-)$,
giving the well-known condition
\begin{equation}\label{SKP}
\cos{ka} = \cos{\k a}+{\a\over 2 \k}\sin{\k a}.
\end{equation}

We can construct a non-Hermitian PT-symmetric version of this
model by taking the coefficients of the delta functions to be pure
imaginary and alternating in sign, and arranging them
symmetrically about the origin:
\beq\label{thetaeq}
V(x) = i\a \sum_n (-1)^n
\delta(x-\half(n+\half)a).
\eeq
The periodicity of the potential is still
$a$. To perform the Floquet analysis it is easier to shift the
origin to $-{1\over 4}a$, so that the two wave functions initially
take the same form as in Eq.~(\ref{Floquet}) and then track their
discontinuities through the delta functions at $x=\half a$ and
$x=a$. The net result is that the Kronig-Penney expression for
$\cos ka$ is replaced by
\begin{equation}\label{iKP}
\cos{ka} = \cos{\k a} + {\a^2\over 2\k^2}\sin^2{\half \k a},
\end{equation}
from which it is immediately apparent that $\cos{ka}$ is strictly
greater than -1, and in particular that there are no antiperiodic
solutions with $k=(2n+1)\pi/a$.

A generalization of Eq.~(\ref{thetaeq}) is the potential
\beq
V(x) = \a \sum_n e^{(-1)^n i\th}\delta(x-\half(n+\half)a),
\eeq
which reduces to Eq.~(\ref{thetaeq}) for $\th=\pi/2$ and to the standard
Kronig-Penney model, with spacing $\half a$, for $\th=0$.
It is then a reasonable question to
ask at what value of $\th$ the antiperiodic solutions disappear:
the answer is perhaps surprising.

Beginning, as before, with the two solutions of
Eq.~(\ref{Floquet}), and tracking the discontinuities of their
derivatives through the (shifted) delta functions at $x=\half a$
and $x=a$, we arrive at the equation
\beq\label{thetaKP}
\cos{ka}
= \cos{\k a}+{\a\over\k}\sin{\k a}\, \cos\th +{\a^2\over
2\k^2}\sin^2{\half\k a},
\eeq
which reduces to (\ref{iKP}) for
$\th=\pi/2$. It implies the following relation for $\cos{ka}+1$:
\beq\label{thetaKPp}
\cos{ka}+1=2\left| \cos{\half\k a} +{\a \over 2
\k}e^{i\th} \sin{\half\k a} \right|^2,
\eeq
which is non-zero as soon as $\th \neq 0,\; m\pi$. Thus the
antiperiodic solution disappears immediately, rather than at some
finite critical angle between 0 and $\pi/2$.

As noted above, whereas for $\th\neq 0$ the repeat distance is $a$, it
reduces to $\half a$ at $\th=0$. Correspondingly (\ref{thetaKP})
can be rewritten as
\beq\label{theta=0}
\cos{\half ka} = \cos{\half\k a}
+{\a\over 2\k}\sin{\half\k a},
\eeq
 which starts off positive,
passes through $\cos{\half ka}=0$ and then becomes negative.
However, the moment $\th$ becomes non-zero this solution
disappears, and by continuity
 $\cos{\half ka}$ must always remain positive, i.e.
\beq\label{theta<>0}
\cos{\half ka}=\left| \cos{\half\k a} +{\a \over 2
\k} e^{i\th}\sin{\half\k a} \right|.
\eeq
The situation is illustrated in Fig.~1, where we have taken
$\th=0.1^{\rm c}$.

\begin{figure}[h]
\centerline{\epsfxsize=3in
\epsffile{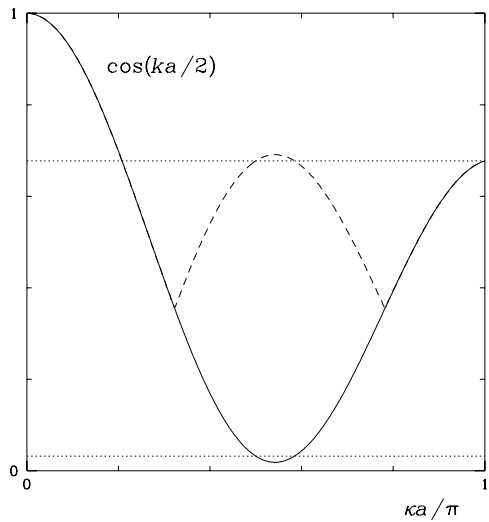}}
\caption{Eq.~(\ref{theta=0}) (solid line) and
Eq.~(\ref{theta<>0}) (dashed line). Where both expressions are
positive the two curves are barely distinguishable.}
\end{figure}

In Fig.~2 we show the resulting band structure for the same value
of $\theta$.
The Brillouin zone boundary is at $k=\pi/a$, in the middle of
the Brillouin zone for $\theta=0$. In contrast to the usual
situation, as exemplified by the dispersion relation for phonons
in a diatomic molecule, where a gap appears at the boundary, here
the energy levels merge together before reaching the boundary, thus
forming one continuous, double-valued band. At the point where the
levels merge the effective mass is zero.

\begin{figure}[h]
\centerline{\epsfxsize=3.9in
\epsffile{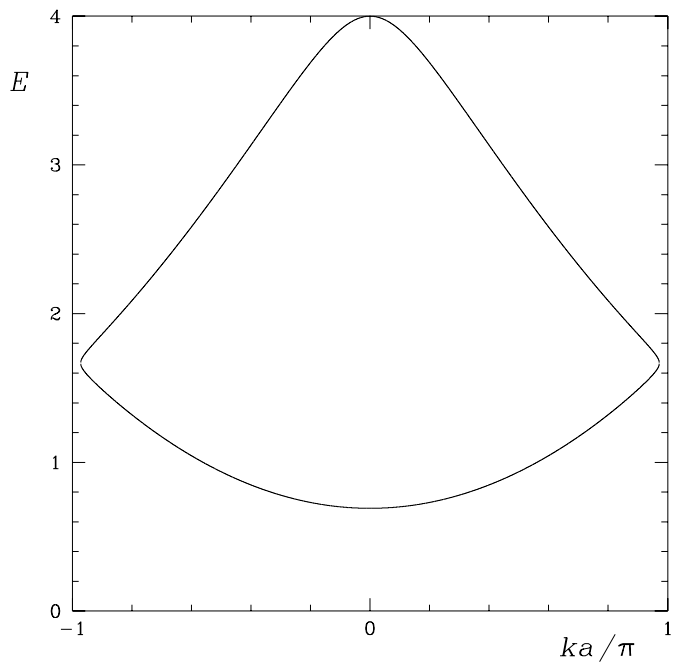}}
\caption{The structure of the lowest band resulting from
Eq.~(\ref{thetaKPp}). Energy in arbitrary units.}
\end{figure}

Consideration of this toy model seems to show that there is
something generic about the dispersion relations for periodic
potentials with $V^*(x)=V(-x)\neq V(x)$. In this case we have an
analytic demonstration that as soon as $\theta\neq 0$ there
is no antiperiodic solution and the energy levels do not reach
the Brillouin zone boundaries. A general proof of this property,
not depending on the specific form of $V$, would be very welcome.

{\bf Acknowledgement}

I should like to thank Prof. C.~M.~Bender for his hospitality at
Washington University, where this work was begun.


\end{document}